# A new efficient Matching method for web services substitution

Jaouad Boutahar[1], Taoufik Rachad[2] and Souhail El Ghazi El Houssaini [1]

[1] Systems, architectures and networks team, EHTP
Casablanca, B.P 8108 Oasis, Morocco

[2] Architectures and systems team, LISER Laboratory, ENSEM
Casablanca , BP 8118 Oasis, Morocco

**Abstract**
The internet is considered as the most extensive market in the world. To keep its gradual reputation, it must confront real problems that result from its distribution and from the diversity of the protocols used to insure communications.
The Web service technology has diminished significantly the effects of distribution and heterogeneity, but there are several problems that weaken their performance (unavailability, load increase of use, high cost of CPU time...). Faced with this situation, we are forced to move in the direction of the substitution of web services. In this context, we propose an effective technique of substitution based on a new method of matching that allows detecting and expressing the matching between the web services pairwise by considering that each of them is ontology. Also, our method performs a discovery of the most similar web service to that to be replaced by using an efficient method of similarity measurement.

## 1. Introduction

Nowadays, the internet has become a huge market for the sale of goods and services, its sophisticated technological support, its very reduced cost with regard to other marketing channels as well as its secure protocols have led to the fact that millions of companies and people use it daily to do their business. This increased use of the Internet has become a necessity to be fully able to benefit from its opportunities. In this respect, new concepts have emerged such as B2B, B2C and C2C.
The diversity of technologies and protocols used on the Internet in addition to its distributed nature has given rise to a remarkable technological heterogeneity. To this effect, the need for collaboration and communication via the Internet has led organizations and stakeholders in the field of internet to propose new techniques and protocols which remove or or at least reduce this heterogeneity. The service concept and especially its implementation web service is one of those technologies that have proved to be very successful.

Web services can be defined as "software applications, loosely coupled with dynamic interaction, identified by a URI, which may interact with other software components whose interfaces and bindings have the capacity to be published, located and invoked via XML and use of common Internet protocols. More specifically, web services are based on the use of three basic protocols: SOAP(Simple Object Access Protocol) for communication and exchange of XML messages, WSDL (Web Services Description Language) for describing the web service and UDDI (Universal Description, Discovery and Integration) for the publication of web service

Due to its simplicity and flexibility, the web service technology has become a necessity to ensure mutual interoperability with collaborators on the internet. Several applications consume the available web services on the internet and as a result their performances depend strongly on those of these web services.
Currently, there are several problems that weaken the web services performances:

- Their low availability or downright their defections: They must therefore be replaced by other similar web services.
- The scalability of their use: They are overstretched; hence the need to find other similar web services to be executed in parallel, to reduce the load of use.
- The high cost of CPU time on the execution: they must, therefore, be replaced by a low cost web services.

Faced with this situation and to improve the performance of web services, we are often forced to move in the direction of their substitution.
The Web service substitution is the domain that deals with methods and techniques that allow replacing a web service by another one that is similar and more efficient. Substitution encounters several problems such as the existence of web services offering the same services, but do not have the same interfaces and the existence of web services that belong to the same business domain, but does not share the same vocabulary.

To provide solutions to these problems, several works such as (owls [1], WSDLs [2] owmo [3], ...) were mainly based on global domain ontologies that define a unified vocabulary. All web services should be semantically linked to the global ontology, which will allow a web service to be replaced by any other web service in an automatic way without being wary of semantic differences. The implementation of such ontologies is a costly operation both in terms of time and effort because of the lack of standard ontologies and the lack of sophisticated tools for generating ontologies. Even if they exist, these ontologies are used to remove the ambiguity between web services at the data level and do not allow to eliminate the ambiguity for operations enchainment at the process level.

In our work, we propose a new method that exploits the web services WSDL interfaces as ontologies and that describe a web service on three levels, namely the data level, the functional level and the behavior level (operations enchainment). Thereafter, unlike methods based on ontologies that describe all the available web services using the terms of an ontology, our method allows to express the correspondence between web services pairwise by considering that each web service is an ontology, something that is much easier and more accurate than using shared ontologies. Before expressing the correspondence between two web services (substituted and substituent), our method performs a discovery of the most similar web services to the web service to replace (substituted) using the technique of calculating similarity between web services as described in [4] and described briefly in this paper.

In Section II of this paper we present our vision in the use of information embedded in a WSDL file. In section III we present the model and the technique used to detect the correspondence between two web services in the context of web service substitution. In section IV we present the implementation of our model. We conclude later in Section V and we present some perspectives.

## 2. Preparation of a WSDL file structure

A WSDL (Web Service Description Language) file is an XML file that follows a standard format for describing a web service. It mainly describes the provided operations and how to access them and describes the data schema used for communication with the web service. Figure 1 bellow illustrates the different elements of a WSDL file.

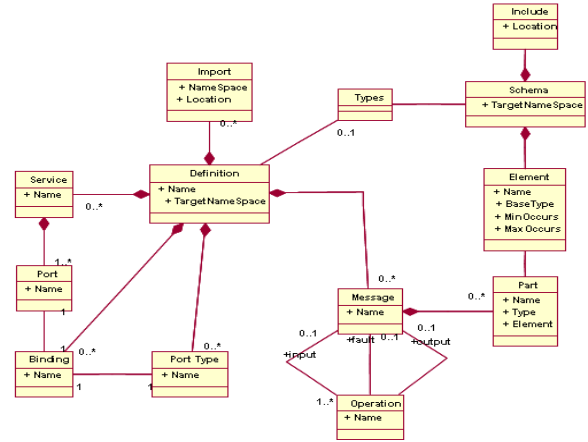

Fig. 1  WSDL (web service description language) model.

In our work we consider that a web service is a set of operations, each of them has an identifier, a given input and output data. Any input or output of a web service operation is considered as a subset of the wsdl data schema that is named element. An element in a data schema has a tree structures. The name of the element is the root of the tree, the internal nodes correspond to the elements that have complex types and the leaves of the tree correspond to elements that have simple types (example in Figure 2).

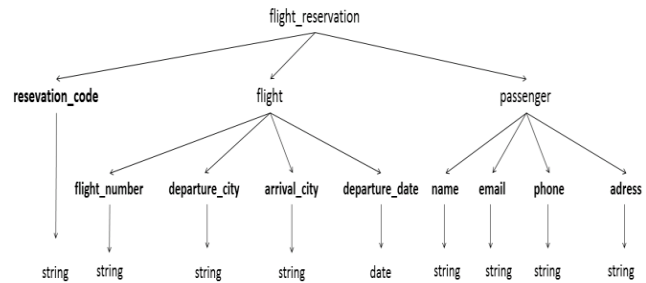

Fig. 2  example of a tree structure of a data set

The tree structure of an element is very complex and complicates the operation of matching. To overcome this problem, we apply to an element a transformation that will provide it with a tree structure with one level (the root directly connected to the leaves). The leaves names will be concatenated with the names of the nodes that connect them with root. In this way, a leaf will represent a whole path in the tree without giving any importance to the tree structure. Thus, a schema element is considered as a set of sentences, each one represents a leaf of the tree. For illustration, an example of a one level tree structure is shown in Figure 3.

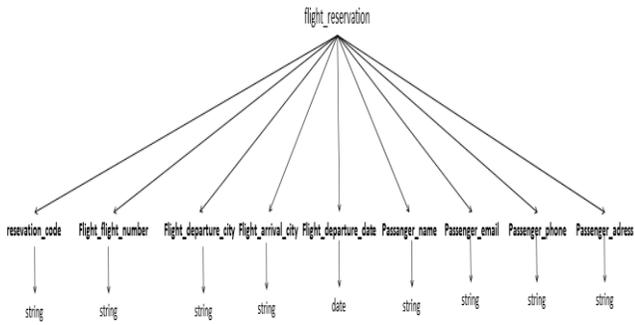

Fig. 3 tree structure with one level

So, we consider that whatever *WS* a web service with a data schema *S*, *WS={op_i}* such that $op_i$ is the $i^{th}$ operation of the web service *WS*.

∀ $op_i$∈*WS*, $op_i$ has an input $IN_i$∈ *P (S)* and an output $OUT_i$∈ *P (S)* such that *P(S)* is the set of parts of the schema *S*. $IN_i$ and $OUT_i$ are two sets of strings that are obtained by applying a data transformation to their elements. Figure 4 represent the used WSDL model for detecting matching between web services.

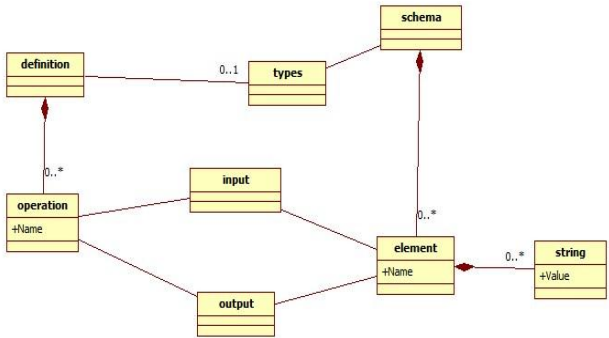

Fig. 4 Used WSDL model

## 3. The proposed model

The proposed model for the realization of the matching between web services (Figure 5) consists of four modules: The module of basic similarity (between strings), the module of similarity between web services, the module of matching between web services and the module of semantic annotation of web services.

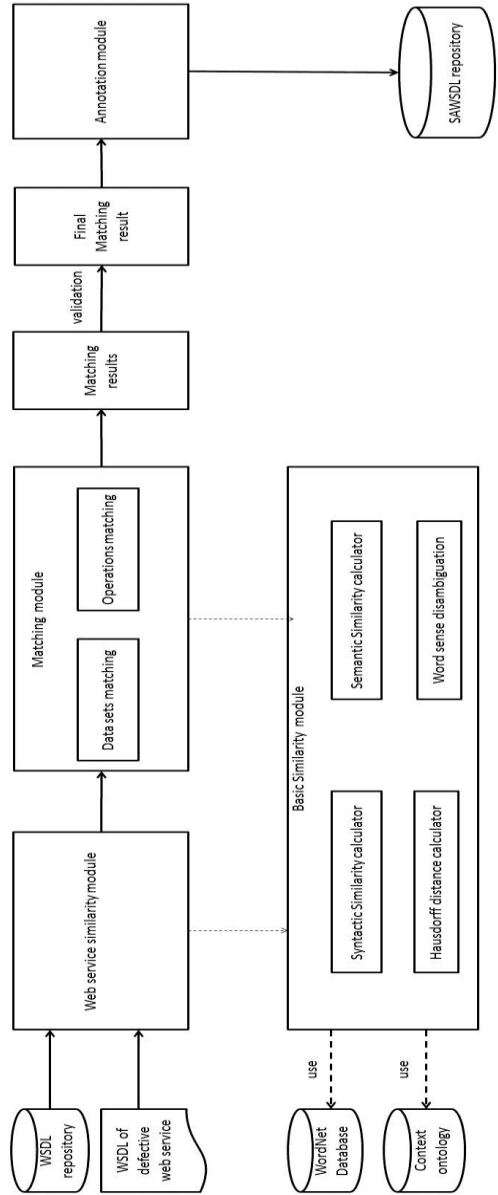

Fig. 5 the proposed model for detecting the matching between web services

The 'basic similarity' module can calculate the similarity between two strings. It consists of four sub-modules, namely the syntactic similarity module, the semantic similarity module, the word sense disambiguation module and the module for calculating the Hausdorff distance.

The 'similarity between web services' module uses the basic similarity module and allow to filters the available web services in the UDDI registry, so as to identify those that can match the web service to replace and those that do not match .

The 'Matching' module uses the basic similarity module and is mainly used to detect the matching between any two web services operations, it consists of two sub-modules, namely the 'Data sets matching' module that can detect the matching between the inputs and between the outputs of operations of any two web services and 'matching operations' module that can detect all possible relations between the operations of any two web service based on the results of the basic similarity module and the data sets matching module.

After validating the results of matching, the 'Semantic Annotation' module will annotate the two WSDL interfaces of substituted and substituent web service, subject of matching, with the matching results, in order to take them into account during the invocation of substituent web service.

### 3.1 Basic similarity module

This module is mainly used to calculate the similarity between two strings and will be used particularly to measure the similarity between any two objects in the WSDL description of a web service, ie two words, two sentences, two inputs, two outputs, two operations and two web services. The operation completed by this module has already been the subject of a previous work [4] which we briefly present in this section.

### 3.1.1 Syntactic similarity Calculator

The syntactic similarity consists of assigning to a pair of strings S1 and S2 a real number, which indicates the degree of syntactic similarity between them. Several algorithms presented in [4] are used to perform this measurement of which the most powerful is Jaro-Winkler [5], [6] which will be retained in this work.

### 3.1.2 Semantic similarity Calculator

The semantic similarity consists in assigning to a pair of words $w_1$ and $w_2$ a real number, which indicates the degree of semantic similarity between them. The similarity measurement is done by comparing the senses of the two words. Thus, any two words are similar (with a certain degree of similarity) semantically if they mean the same thing (synonyms). They are used in the same way or inherit the same type, they are used in the same context or if one is a type of the other. To measure the semantic similarity between words, we will need a lexical hierarchy such as WordNet [7].

According to the study that we performed in [4], the Wu-Palmer algorithm [8] is the most effective and the most efficient in the calculation of semantic similarity.

### 3.1.3 Word sense disambiguation

The semantic similarity measurement between two words refers to the measurement of similarity between the meanings of the two words. All algorithms for measuring semantic similarity, consider the most common meanings or meanings that offer the greatest similarity during the comparison process. But the meaning of a word changes according to the context in which the word appears. That is why we must extract the exact meanings of words before tackling the similarity measurement. Word sense disambiguation is the scientific term that has been attributed to the process of searching for the exact meaning of a word in a specific context. Adapted Lesk algorithm described in [9, 10] is the algorithm adopted to lift the ambiguity of word sense in a given context. We proposed an implementation of this algorithm in [4].

### 3.1.4 Haudorff distance caclulator

The Hausdorff distance is used to measure the distance between two sets of points. In [11] the authors conclude that the modified Hausdorff distance has the highest performance in the similarity measure between two sets of points.

Assuming that d is any distance, the modified Hausdorff distance between two sets of points S and T is defined by the relation:

$$HD(S,T) = \max\left\{ \frac{1}{|S|}\sum_{p \in S} \min_{q \in T}\{d(p,q)\} \quad , \quad \frac{1}{|T|}\sum_{q \in T} \min_{p \in S}\{d(p,q)\} \right\} \quad (1)$$

Hausdorff distance will be used later to measure the distance between any two sets of objects (words, sentences, operations ...).

### 3.1.5 Similarity between two strings

A string is a set of words separated by connectors. Consequently, the similarity between two strings will be the similarity between the sets of words that represents them. Let $S_1$ and $S_2$ two strings such that $S1=\{w_0,w_1,......w_n\}$ and $S2=\{w'_0,w'_1,.....w'_m\}$. Let $MW$ a similarity matrix such that $\forall (w_i,w_j') \in S_1 \times S_2$ $MW(i,j)=WordSimilarity(w_i,w_j',context)$. the $WordSimilarity$ function implemented in [4] return the similarity between any two words in a very specific context, it tries firstly to measure the semantic similarity between two words using the WuPalme algorithm, if one of the two words do not exist in WordNet then it returns a syntactic similarity using the JaroWinkler algorithm.

The similarity between two strings $S_1$ and $S_2$ will be the Hausdorff distance between the sets of words of $S_1$ and $S_2$, this distance is defined by the relation:

$$SentenceSimilarity(S1,S2) = HD(S1,S2) =$$
$$= \max\left\{ \frac{1}{n}\sum_{i=0}^{n-1} \min_{0 \le j < m}\{MW(i,j)\}, \frac{1}{m}\sum_{j=0}^{m-1} \min_{0 \le i < n}\{MW(i,j)\} \right\} \quad (2)$$

### 3.2 Similarity between web services module

### 3.2.1 Similarity between two data sets

The measurement of the similarity $SetSimilarity(DS_1,DS_2)$ between two data sets $DS_1$ and $DS_2$ (that will be either inputs or outputs of web services operations) will be done by measuring the Hausdorff distance between $DS_1$ and $DS_2$. By considering that $DS_1$ and $DS_2$ are two sets of sentences, the Hausdorff distance uses a similarity matrix $MS$ such that $\forall (S_i,S'_j) \in DS_1 \times DS_2 \; MDS(i,j)=SentenceSim(S_i,S'_j)$.

$$SetSimilarity(DS1,DS2) = HD(DS1,DS2) = \max\{ \frac{1}{n}\sum_{i=0}^{n-1}\min_{0\leq j<m}\{MDS(i,j)\}, \frac{1}{m}\sum_{j=0}^{m-1}\min_{0\leq i<n}\{MDS(i,j)\} \} \quad (3)$$

### 3.2.2 Similarity between two operations

Let f and g be two web service operations, the similarity between f and g is the sum of the similarity between their inputs sets, the similarity between their outputs sets and the similarity between their names:
$OperationSimilarity(f,g)=p1*SetSimilarity(D,D')+p2*SetSimilarity(A,A')+p3*SentenceSimilarity(f.name,g.name)/(P1+P2+P3)$

In the calculation, we used a weighting that determines the order of importance of each similarity function. In the measures that we have taken, it was considered that $p_1 = 1$, $p_2 = 1$ and $p_3 = 2$.

### 3.2.3 Similarity between two web services

In our work, a Web service is considered as a set of operations. The similarity between two web services $WS_1$ and $WS_2$ will be the Hausdorff distance between the two sets representing operations. Hausdorff distance uses a similarity matrix MWS such that $\forall (op_i,op_j') \in WS_1 \times WS_2 \; MWS(i,j)= OperationSimilarity(op_i,op_j')$.

$$ServiceSimilarity(WS1,WS2) = HD(WS1,WS2) = \max\{ \frac{1}{n}\sum_{i=0}^{n-1}\min_{0\leq j<m}\{MWS(i,j)\}, \frac{1}{m}\sum_{j=0}^{m-1}\min_{0\leq i<n}\{MWS(i,j)\} \} \quad (4)$$

### 3.3 Matching module

After measuring the similarity between the web service to be replaced and the available web services, only the web service that is most similar will be retained. The next step is to detect the correspondence between the operations of the two web services (substituted and substituent).
Let $WS_1$ and $WS_2$ two web services for which we want to carry out the matching. Let $S1$ and $S2$ their data schemas and $OP1$ and $OP2$ two sets of operations such that: $OP_1= \{op_{1i}\}$ is the set of $WS_1$ operations and $OP_2= \{op_{2i}\}$ is the set of $WS_2$ operations.
$\forall op_{1i} \in OP_1$ and $\forall op_{2j} \in OP_2$ we must determine the relation $R(op_{1i}, op_{2j})$ connecting the two operations $op_{1i}$ and $op_{2j}$ such that $R \in$ {equality restriction corestriction, prolongation, difference, intersection). The determination of the relation that links the two operations $op1i$ and $op2j$ depends on the relations that connect their input and output. So $\forall op_{1i} \in OP_1$ and $\forall op_{2j} \in OP_2$, we must determine the relations $R_1(IN_{1i},IN_{2j})$ and $R_2(OUT_{1i},OUT_{2j})$ as $IN_{1i}$ and $IN_{2j}$ are respectively the inputs of $op_{1i}$ and $op_{2j}$, as well as $OUT_{1i}$ and $OUT_{2j}$ are respectively their outputs with $R_1, R_2 \in \{=, \neq, \subset, \supset, \cap\}$. In Table 1, we illustrate all possible cases of connecting any two operations.

We define the relations that can connect two operations as follows:
- Restriction: two operations $op_{1i}$ and $op_{2j}$ are connected by a relation of restriction if $OUT_{1i}= OUT_{2j}$ and ( $IN_{1i} \subset IN_{2j}$ or $IN_{2j} \subset IN_{1i}$ ).
- Corestriction : : two operations $op_{1i}$ and $op_{2j}$ are connected by a relation of corestriction if $IN_{1i}= IN_{2j}$ and ( $OUT_{1i} \subset OUT_{2j}$ or $OUT_{1i} \subset OUT_{2j}$ )
- Prolongation : two operations $op_{1i}$ and $op_{2j}$ are connected by a relation of prolongation if ( $IN_{1i} \subset IN_{2j}$ and $OUT_{1i} \subset OUT_{2j}$ ) or ( $IN_{2j} \subset IN_{1i}$ and $OUT_{2j} \subset OUT_{1i}$ )
- Equality: two operations $op_{1i}$ and $op_{2j}$ are connected by a relation of equality if $IN_{1i}=IN_{2j}$ and $OUT_{1i}= OUT_{2j}$
- Difference : two operations $op_{1i}$ and $op_{2j}$ are connected by a relation of difference if $IN_{1i} \neq IN_{2j}$ and/or $OUT_{1i} \neq OUT_{2j}$
- Intersection: two operations $op_{1i}$ and $op_{2j}$ are connected by a relation of intersection in all remaining cases.

To determine the relation between the data sets $IN_{1i}$ and $IN_{2j}$ on the one hand and between $OUT_{1i}$ and $OUT_{2j}$ on the other hand, we consider them as sets of sentences as shown above in examples of figure2 and figure 3.
Thus, for each two data sets $E$ and $E'$ such that $E=\{S_i\}$ and $E'=\{S'j\}$ we use a similarity matrix $MS$ such that $\forall (S_i,S'_j) \in E \times E' : MS(i,j)=V_{ij}=SentenceSimilarity(S_i,S'j)$.
The SentenceSimilarity function returns the similarity between the two strings $S_i$ and $S'_j$.
Using the rules below, we can extract the binary relations that can connect any two data sets E and E ':

- $E \subset E' \Leftrightarrow \forall i \in [0,n], \exists j / v_{ij} > threshold$
- $E \equiv E' \Leftrightarrow E \subset E' \; and \; E' \subset E$
- $E \cap E' \Leftrightarrow \exists (i,j) \in [0,n] \times [0,m] / v_{ij} > threshold$
- $E \neq E' \Leftrightarrow \forall (i,j) \in [0,n] \times [0,m] \; v_{ij} \leq threshold$

In the implementation of our model we considered that the threshold is 0.5.

Table 1: the possible cases of connecting two operations

| R1(IN1i, IN2j) | R2(OUT1i,OUT2j) | R(op1i, op2j) | | | | | |
|---|---|---|---|---|---|---|---|
| | | equality | restriction | coRestriction | prolongation | intersection | difference |
| = | = | * | | | | | |
| = | ⊂ | | | * | | | |
| = | ⊃ | | | * | | | |
| = | ∩ | | | | | * | |
| = | ≠ | | | | | | * |
| ⊂ | = | | * | | | | |
| ⊂ | ⊂ | | | | * | | |
| ⊂ | ⊃ | | | | | * | |
| ⊂ | ∩ | | | | | * | |
| ⊂ | ≠ | | | | | | * |
| ⊃ | = | | * | | | | |
| ⊃ | ⊂ | | | | | * | |
| ⊃ | ⊃ | | | | * | | |
| ⊃ | ∩ | | | | | * | |
| ⊃ | ≠ | | | | | | * |
| ∩ | = | | | | | * | |
| ∩ | ⊂ | | | | | * | |
| ∩ | ⊃ | | | | | * | |
| ∩ | ∩ | | | | | * | |
| ∩ | ≠ | | | | | | * |
| ≠ | = | | | | | | * |
| ≠ | ⊂ | | | | | | * |
| ≠ | ⊃ | | | | | | * |
| ≠ | ∩ | | | | | | * |
| ≠ | ≠ | | | | | | * |

Thereafter, an administrator must assess the table of correspondence and must determine the correct matching between the operations of both substituted and substituent web services. He must take into account the priority of relations of correspondence (1-equality, 2-co-restriction, 3-restriction, 4-prolongation, 5-and 6-intersection difference) and the similarity rank.

The matching module returns the result of its treatment in the form of a table of correspondence between the operations of the web service to be replaced and the web service chosen during the operation of similarity. For every pair of operations op1 (of the web service to be replaced) and op2 (of the similar web service) the correspondence table gives the relation that connects them and the similarity rank. Table 2 illustrates this correspondence table obtained by performing the matching between two web services in the field of weather.

Table 2: example of correspondence table

| | GetWeather | getCity |
|---|---|---|
| GetWeather | Equality(0.95) | Restriction (0.71) |
| GetcitiesByconty | Intersection(0.59) | Restrinction0.8) |

An operation of the web service to be replaced can correspond to one or more operations on the similar web service, so that during the expression of the matching the administrator can use the AND and OR operators.

After that the administrator expresses the correspondence between the operations of the web service to be replaced and the operations of similar web service, the next step is to express the correct match between their inputs and their outputs.

Let $op_{1i}$ an operation of the web service to be replaced and $op_{2j}$ the corresponding operation on the similar web service, at invoking stage, the operation $op_{1i}$ must be replaced by the operation $op_{2j}$, so, the administrator must express the inputs of $op_{2j}$ based on inputs of $op_{1i}$ to realize data transformations during the invocation, and for that, arithmetic operators (+, -, * and /) and concatenation operator can be used. Eventually, a correspondence table between the inputs of the two operations (see implementation section) can be used. This operation is repeated as many times as the operations of the similar web service occur in matching with an operation of the web service to be replaced.

For outputs, the match must be performed in reverse, ie that the output of the operation of the web service to be replaced must be expressed in terms of outputs of all similar operations.

3.4 Annotation module

To express the found matching, we use the SAWSDL language (Semantic Annotations for WSDL and XML Schema) [12] which is a w3c standard that supports the semantic description of a web service in a simple way. It adds a semantic description to the various elements of a WSDL file by using XML annotations. The annotation refers to a semantic model such as ontology.

SAWSDL does not specify a specific language to represent the semantic model, but any XML file can be used.

SAWSDL uses two ways to annotate WSDL elements, the first one is the annotation by modelReference and the second is the annotation by SchemaMapping.

SchemaMapping is used to bind a WSDL schema element to ontology by using a data transformation. The transformation definition can be defined in two different ways, a liftingSchemaMapping definition or loweringSchemaMapping definition [12].

The modelReference annotation can be used with any element of the wsdl description of a web service, namely wsdl: interface, wsdl: operation, wsdl: fault, xs: element, xs: complexType, xs: simpleType and xs: attribute (example in Figure 6). In our work we are interested only in the modelReference annotation.

In our work we use a WSDL description as a semantic model. The found matching will be expressed by using modelReferences in the wsdl of the substituted web service and the WSDL of the substituent web service.

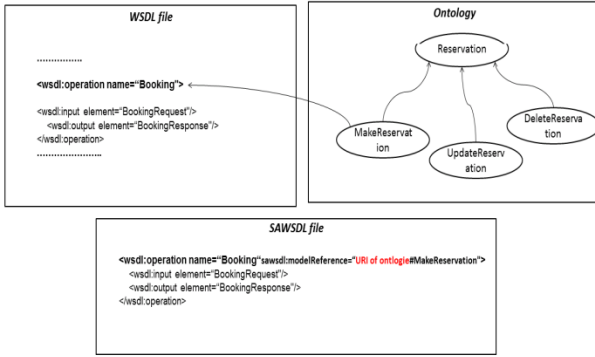

Fig. 6 annotation example with modelRefence

## 4. Implementation

In this part, we implemented a scenario where we are trying to replace a defective web Service by another that is better, the latter is extracted from a set of web services offering similar services and saved in an UDDI registry. We implemented our solution to allow performing the substitution in the best conditions with respect to the cost, time and performance.

At first, the application request the WSDL link of the web service to be replaced and the link of the UDDI registry that contains the web services that may be similar to the web service to be replaced (Figure 7).

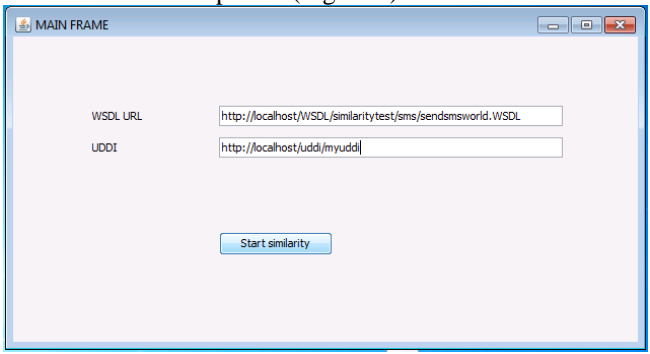

Fig. 7 the main window

Then, the application calculates the similarity between the web service to be replaced and all web services available in the UDDI and returns the result as a list that contains the most similar web services sorted in descending order of similarity (Figure 8) . The user chooses a web service and then asks for the application to start the process of matching.

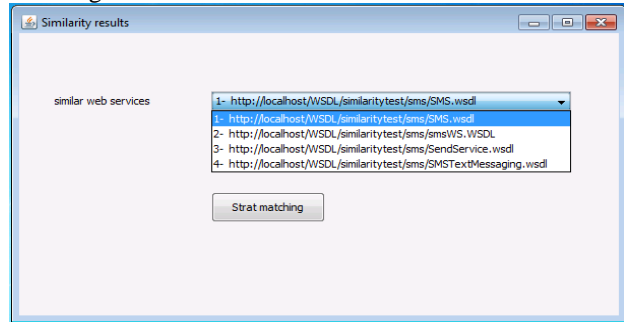

Fig. 8 similarity result

The application displays all possible connections between the operations of the web service to be replaced and the operations of the chosen web service (Figure 9) and the user expresses later the exact match in the annotation area of operations by using the AND and OR operators .

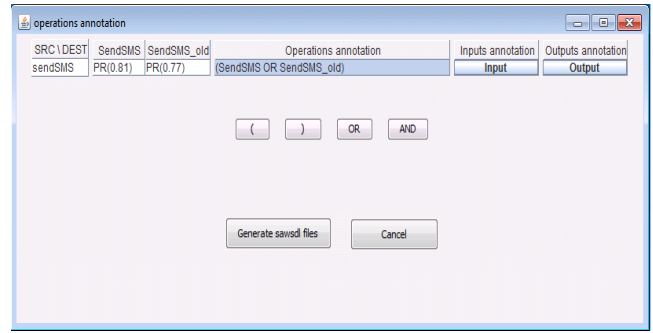

Fig. 9 operations matching and annotation

Thereafter, the user expresses the matching between the inputs and between the outputs of operations involved in the annotation of operations of the web service to be replaced (Figure 10 and Figure 11).

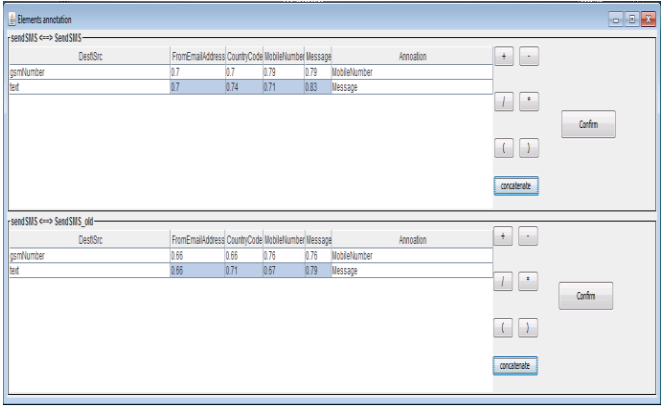

Fig. 10 inputs matching and annotation

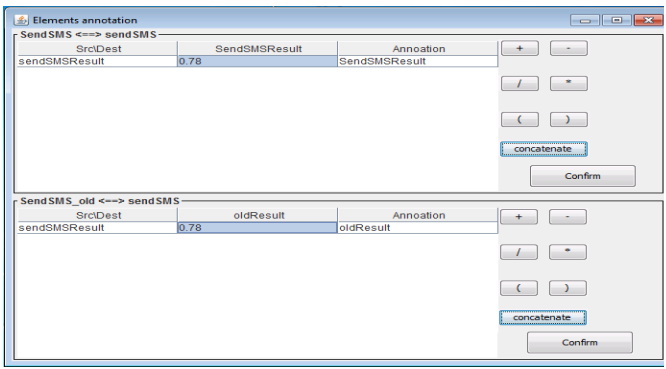

Fig. 11 outputs matching and annotation

Once the user completes the matching and confirms, two WSDLfiles annotated with SAWSDL are generated. Subsequently, the generated files will be used in a transparent manner, , during the invocation process of a web service in the operating environment.

## 4. Conclusion

In this paper, we presented a realistic, simple and effective approach which allows to realize the substitution of web services via two steps: the first step of discovery and similarity calculation that returns web services which are similar to the web service to replace, sorted in ascending order of their similarity rank; the second step of matching for specifying the correspondence between the substituted and substituent web services.
We have developed a tool that implements the proposed model, which supports users in the completion of matching.
We have also tested our approach in several domains to check its performance, and we found out that our method is not complicated and more precise in comparison with other methods which are very complex and not accurate.

## References

[1]. Martin, D., Burstein, M., Mcdermott, D., Mcilraith, S., Paolucci, M., Sycara, K., ... & Srinivasan, N. (2007). Bringing semantics to web services with OWL-S.World Wide Web, 10(3), 243-277.
[2]. Akkiraju, R., Farrell, J., Miller, J., Nagarajan, M., Schmidt, M., Sheth, A., & Verma, K. (2005). Web Service Semantics-WSDLS. Technical note, April 2005.
[3]. De Bruijn, J., Bussler, C., Domingue, J., Fensel, D., Hepp, M., Kifer, M., ... & Stollberg, M. (2005). Web service modeling ontology (wsmo). Interface, 5, 1.
[4]. Rachad, T., Boutahar, J., & Elghazi, S. (2014). A new efficient method for calculating similarity between web services. IJACSA, 5(8), 60-67.
[5]. Winkler, W. E., "The state of record linkage and current research problems ", Statistics of Income Division, Internal Revenue Service Publication R99/04, 1999.
[6]. Winkler, W. E., " Overview of Record Linkage and Current Research Directions ", Research Report Series, RRS, 2006
[7]. Christiane Fellbaum , "WordNet: An Electronic Lexical DatabaseReferences", Ed. Cambridge: MIT Press, 1998.
[8]. Wu, Z., & Palmer, M. (1994, June). Verbs semantics and lexical selection. InProceedings of the 32nd annual meeting on Association for Computational Linguistics (pp. 133-138). Association for Computational Linguistics.
[9]. Leacock, M Chodorow, "Combining local context and WordNet similarity for word sense identificationC" "WordNet: An Electronic Lexical DatabaseReferences",  Ed. Cambridge: MIT Press, 1998.
[10]. S Banerjee, T Pedersen , "An adapted Lesk algorithm for word sense disambiguation using WordNet", Proceedings of the Third International Conference on Computational Linguistics and Intelligent Text Processing, Pages 136-145, Springer-Verlag London, UK  (2002).
[11]. Marie-Pierre Dubuisson and Anil K. Jain, "A modified Hausdorff distance for object matching", In Proceedings of 12th International Conference on Pattern Recognition, pages 566-568, Jerusalem, Israël, october 1994.
[12]. Kopecky, J., Vitvar, T., Bournez, C., & Farrell, J. (2007). Sawsdl: Semantic annotations for wsdl and xml schema. Internet Computing, IEEE, 11(6), 60-67.

**Jaouad Boutahar** received the PhD in  computer science from the ENPC  in 2004, Paris, France.  He is member of  the systems, architectures and  networks team,   of the EHTP, Casablanca, Morocco. His actual main research interest concern System architectures, web services, Multi Agents Systems and  mediation systems. Since 2004 he is a full professor for computer sciences at the EHTP, Casablanca.

**Taoufik Rachad** Was born in Casablanca in 1983, he obtained in 2006 the Bachelor's degree in Computer Science from the Hassan2 University of Casablanca, in 2007 he obtained a degree in computer science from the higher Normal School and later  in 2009  he obtained a Master's degree in Computer Science from the National Institute of Posts and Telecommunications.
He is currently a PhD student in the higher school of electricity and mechanical in Casablanca and is interested in research in the fields of intelligent agents, web services, semantic web and SOA.
Mr. Rachad  now is teacher of Computer Science in preparatory classes for schools of engineers.

**Souhail El Ghazi El Houssaini** received the PhD in computer science from the AIX-MarseilleII University in 1986, Marseille, France. He is responsible of the system architecture and networks team of the EHTP, Casablanca, Morocco. His actual main research interest concern System architectures, web services, Multi Agents Systems and mediation systems. Since 1987 he is a full professor for computer